\documentclass[%
 reprint,
 amsmath,amssymb,
 aps,
]{revtex4-2}

\usepackage{graphicx}
\usepackage{dcolumn}
\usepackage{bm}
\usepackage{dsfont}
\usepackage{comment}

\usepackage{xcolor}

\usepackage{xspace}
\newcommand{\MBT}{$\text{MnBi}_2\text{Te}_4$\xspace}

\usepackage[colorlinks = true,
            linkcolor = blue,
            urlcolor  = blue,
            citecolor = blue,
            anchorcolor = blue]{hyperref}

\begin{document}

\preprint{APS/123-QED}

\title{Exciton collective modes in a bilayer of axion insulator $\text{MnBi}_2 \text{Te}_4$}

\author{Olivia Liebman}
 \email{oliebman@ucla.edu}
\author{Jonathan B. Curtis}%
\affiliation{%
 College of Letters and Science, University of California, Los Angeles
}%
\author{Emily Been}%
\affiliation{%
 College of Letters and Science, University of California, Los Angeles
}%
\author{Prineha Narang}
 \email{prineha@ucla.edu}
\affiliation{%
College of Letters and Science, University of California, Los Angeles
}%

\date{\today}

\begin{abstract}
We investigate the emergence of an exciton condensate and associated collective modes in a bilayer configuration of $\text{MnBi}_2\text{Te}_4$, an antiferromagnetic topological insulator and van der Waals material, recognized for hosting axion physics. 
Utilizing a minimal low-energy Hamiltonian for the two layer system which is gapped by the intrinsic Néel order, we first employ mean-field theory to establish the conditions for exciton condensation. 
Our analysis identifies a nonzero, spin-singlet exciton order parameter which is tuned by external displacement field, temperature, and Coulomb attraction. 
Beyond the mean-field, we explore collective mode fluctuations in the uncondensed phase via many-body perturbation theory and the random phase approximation. From this, we derive the exciton spectral function which allows for a direct comparison between theoretical prediction and experimental observation. We detail how the softening of the collective mode peak is a function of the competition between interlayer detuning and thermal fluctuations. 
This work elucidates how the unique topological and magnetic environment of $\text{MnBi}_2\text{Te}_4$ offers a tunable platform for the realization and manipulation of exciton condensates and the corresponding collective excitations. Our findings contribute to understanding the interplay of topology and bosonic condensates, which could inspire application in optically accessing topological properties, dissipationless transport, and gate-tunable optoelectronics.

\end{abstract}

\maketitle


\section{Introduction}
Excitons, defined as bosonic composite particles composed of bound electron-hole pairs, are crucial for semiconductor optics and hold significant promise for new functionalities in optoelectronics, driving substantial recent research interest ~\cite{Basov.2016, Mak.2016, Wang.2023, Jankowski.2025, Esteve.2024}. The long-standing quest to achieve Bose-Einstein Condensation (BEC) of excitons
opens a pathway to exploring coherent bosonic states and emergent quantum phenomena. 
A number of systems have realized  exciton condensates, such as: quantum Hall bilayers~\cite{Eisenstein2004, Liu.2017}, cavity exciton-polariton condensates out of equilibrium~\cite{Kasprzak.2006}, and more recently some moir{\'e} materials~\cite{Xiong.2023} in the broader class of doped transistion metal dichalcogenides~\cite{Remez.2022}.

Within this exciting landscape, \MBT, stands out as a unique platform due to its classification as an intrinsic antiferromagnetic (AFM) topological insulator and its predicted capacity to host axion physics, where the axion field is denoted by $\theta$. The static axion field takes the quantized value $\theta = \pi$ ~\cite{Li2022, Yao2008, Zhang2019}, and is symmetry-protected by 
inversion $\mathcal{P}$ and magnetic-crystalline symmetry $\mathcal{S} = \mathcal{T}_{\tau_{1/2}}$ where $\mathcal{T}$ is the time-reversal symmetry (TRS) operator and $\tau_{1/2}$ is the half-translation operator~\cite{Zhang.2020, Li.2010}.   
However, long-range AFM order being established in the system corresponds to a staggered Zeeman effect and resultant broken TRS. Fluctuations in this state promotes $\theta$ to a dynamical quantity and allows the axion field to take continuous values between 0 and $\pi$. 
Here the dynamical axion collective mode is understood as the AFM spin wave \cite{Li.2010}. 

\MBT is expected to host rich physics, with even-layer \MBT understood as an axion insulator, odd-layer \MBT as a Chern insulator, and bulk \MBT realizing type-I or -II Weyl semimetal phases in the presence of an external magnetic field, dependent on the orientation of the field~\cite{Zhang2019}.
Its unique layered, van der Waals (vdW) structure, combined with predictions for a tunable dynamical magnetoelectric effect, further underscores its significance for advanced materials research.

\begin{figure*}[t]
\includegraphics[page=3,width=0.8\textwidth,trim={0 150 0 0},clip]{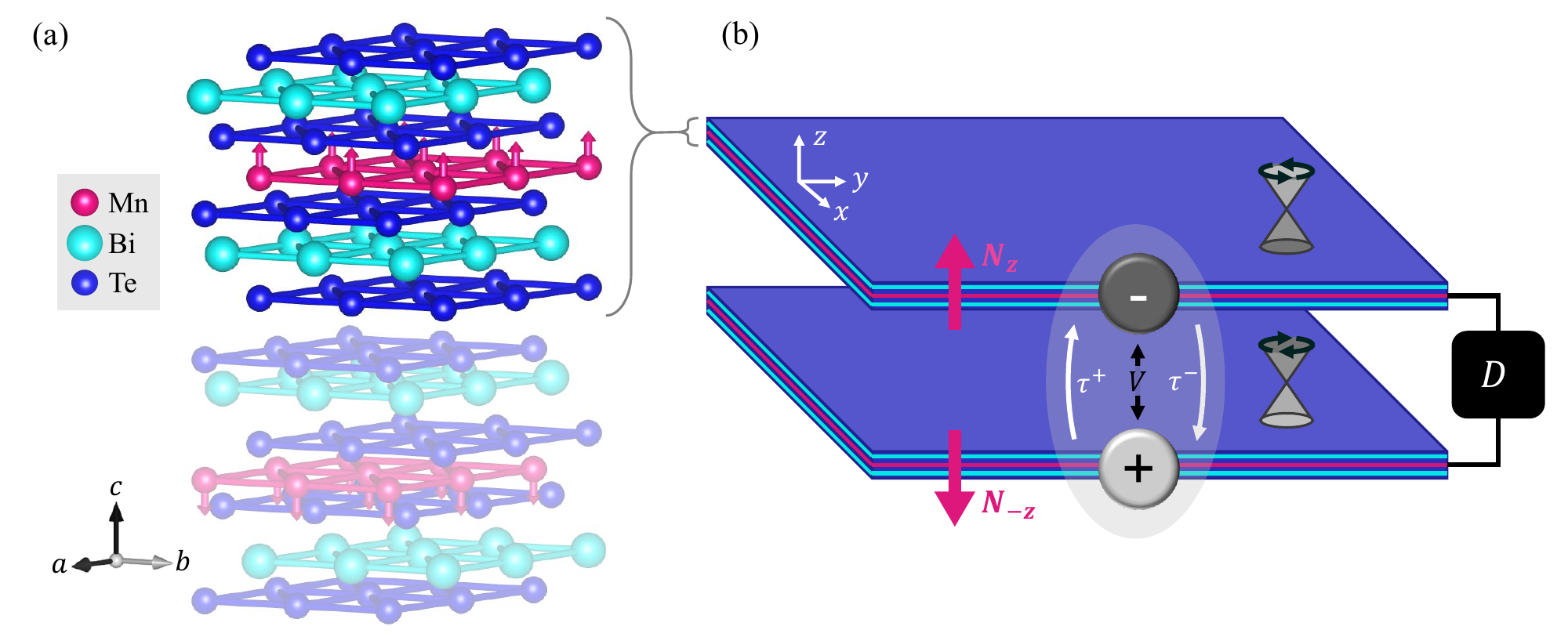}
\caption{\label{fig:MBT_crystal_structure_schematic} Crystal structure of \MBT with space group $R\bar{3}m{\rm :}H$ with (a) showing the isotropic view. A single septuple layer is highlighted, with the AFM magnetic moments of Mn atoms denoted with vectors.
(b) Schematic of the configuration of the two layers of $\text{MnBi}_2\text{Te}_4$, where each sheet indicates a single septuple layer of material. The oppositely spin-polarized Dirac cones on each layer illustrate that this material is a topological insulator. The two layers are connected by an external displacement field, D. The AFM ordering on each layer is indicated by Néel vector, $N_z$, which points along the $z$-direction. The layer operators are denoted as $\tau^{\pm}$ which correspond to taking an electron from the lower layer to the upper layer, and vice versa. The electrostatic Coulomb interaction, $V$, arises due to chemical potential imbalance between the upper and lower layers, which is driven by the displacement field. 
}
\end{figure*}

Here, we show that \MBT may harbor even more exotic physics, by demonstrating how bilayer excitons can form in a two-layer configuration of \MBT. Moreover, under certain conditions, we show how these excitons may condense to form a BEC of excitons. In general, bilayer excitons also have the advantage of more control over excitonic properties and new physics due to the tunability of the distance between the layers, applied tunable bias across the layers, and in some circumstances a tunable twist angle as in twisted bilayer graphene~\cite{Patel.2019, Kim.2024}. 
All this motivates our selection of the bilayer \MBT to study the interplay between topology and exciton order, and how the exciton spectrum may be used to detect nontrivial topology.


Although the physics of excitons and their condensation have garnered significant attention, particularly in systems with extended lifetimes ~\cite{Yao2008,Qi2025}, and axion insulators such as \MBT are recognized for their robust topological properties ~\cite{Zhang2019}, the interplay between exciton condensation and the unique topological-magnetic environment of materials such as \MBT remains largely unexplored. Existing research has predicted spin-dependent exciton transport and vortex formation in exciton condensates due to Berry phase effects, and provided compelling evidence for exciton condensation in bilayer electron systems, even under conditions of electron-hole imbalance leading to TRS-breaking excitonic topological orders ~\cite{Eisenstein2004, Wang2023}. Separately, the concept of a dynamical axion field, where magnetic fluctuations are coupled with electromagnetic fields, has been proposed in topological magnetic insulators, offering new avenues for detection ~\cite{Li2010, Zhang2019, Liebman, Qiu.2025}. However, a detailed theoretical investigation into the formation and collective dynamics of excitons within the complex intrinsic antiferromagnetic and axion-insulator phases of \MBT, especially in a bilayer configuration, is currently missing. Understanding how the Néel order of the material and an external displacement field modulate exciton properties could unveil new functionalities.

\begin{figure*}[t]
\includegraphics[page=6,width=\textwidth,trim={0 20 0 0},clip]
{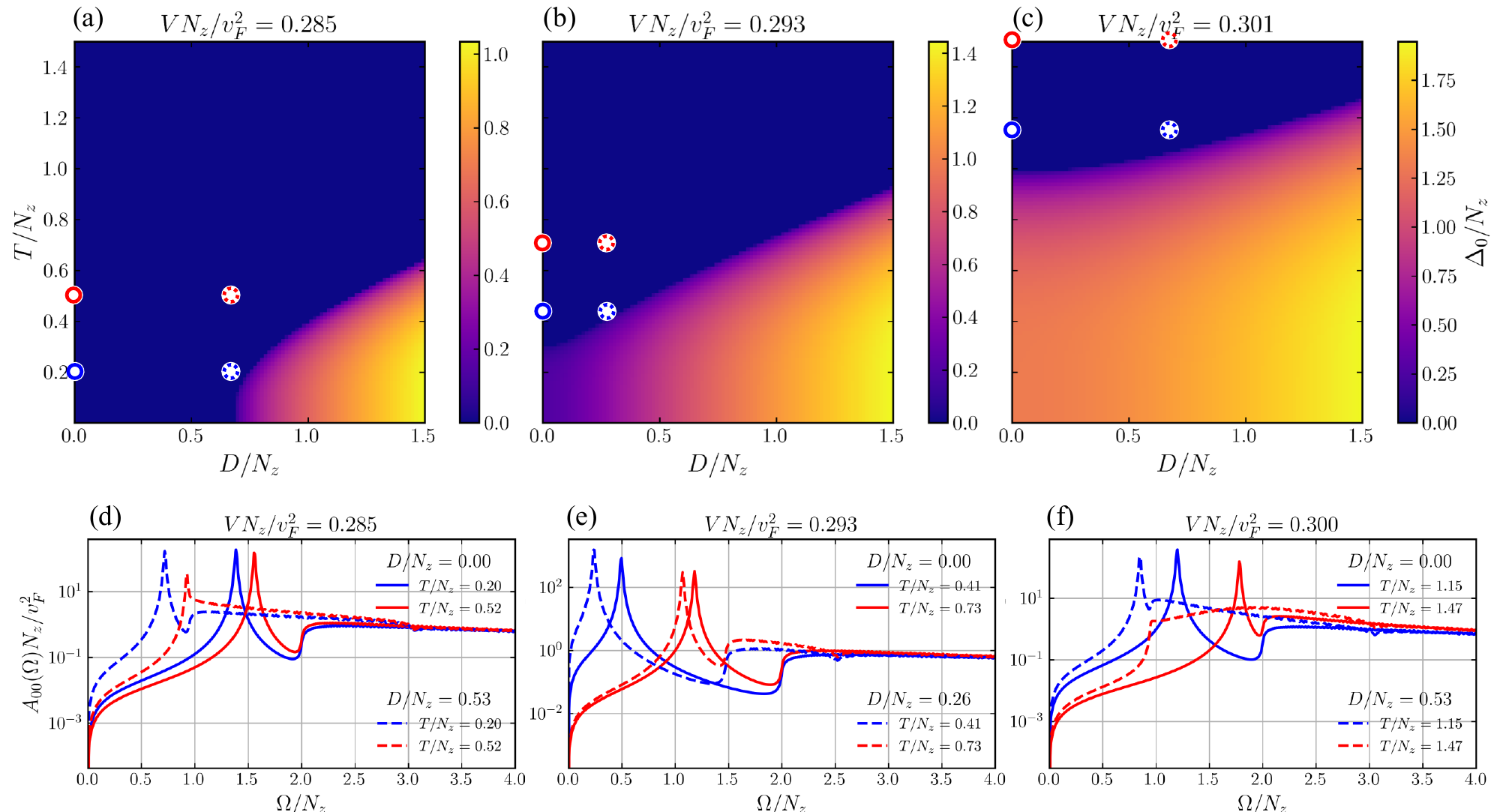}
\caption{\label{fig:PhaseDiagram_SpectralFunction} Upper panel (a)-(c): Mean-field phase diagram of interlayer exciton order parameter $\Delta_0$: 
    color encodes \(|\Delta_0|/N_z\) obtained by solving the self-consistent gap equations derived from the Hubbard--Stratonovich decoupling of the interlayer Coulomb term from Eqs.~\eqref{eq:S_eff_minimize}-\eqref{eq:G_action}.
    Axes are the dimensionless displacement field $D / N_z$ and temperature $T / N_z$; panels (a)-(c) compare the interaction strengths $V N_z / v_F^2 = 0.285,0.293,$ and $0.301$.
    At moderate values of $V N_z / v_F^2$, a finite \(\Delta_0\) appears only once \(D\) exceeds a threshold \(D_c(T)\) which creates a population imbalance of electrons (upper layer) and holes (lower layer).
    The critical line \(T_c(D)\) rises with both $D$ and $V$ as the interlayer particle-hole susceptibility grows, while the overall energy scale of \(\Delta_0\) and \(T_c\) is set by the N\'eel order \(N_z\), which fixes the Dirac gap and available phase space for pairing.
    All energies are normalized by \(N_z\), and \(v_F\) is the Fermi velocity. 
    Lower panel (e)-(g): exciton spectral function $A_{00}(\Omega, \boldsymbol{q}=0)$, in the uncondensed phase ($T>T_c$).
  Each subfigure shows two values of displacement field $D$ (solid vs.\ dashed curves) at fixed interaction strengths (a) $V N_z/v_F^2{=}0.285$, (b) $0.293$, (c) $0.300$, and compares two temperatures (red vs.\ blue curves).
  Note that as we consider higher D values, we must also increase the temperature so that we stay outside of the condensate dome shown in the phase diagram above each subfigure. The displacement field and temperature for each subfigure are indicated on the phase diagram plot immediately above it.  
}
\end{figure*}

Here, we investigate the emergence of an exciton condensate and its associated collective modes in bilayer \MBT, establishing it as a novel and tunable platform for studying the competition between magnetic versus excitonic order. 
Our study develops a minimal low-energy Hamiltonian and employs mean-field theory to delineate the conditions necessary for exciton condensation in this system. Due to the simplicity of the model it is physically motivated to consider only the exciton order parameter with $S=0$ no spin-polarization. However, the framework is sufficiently generic to allow for spin-polarization in future work. The order parameter is critically dependent on factors such as an external displacement field and the intrinsic Néel order of \MBT. 
Furthermore, by extending our analysis beyond mean-field theory to include many-body perturbation theory and the random phase approximation, we characterize the spectral function of these collective modes, demonstrating their emergence from layer-coupling and the gapped Dirac cones induced by the Néel order. These findings highlight how \MBT's unique topological and magnetic properties provide an unprecedented tunable platform for realizing and manipulating exciton condensates and their collective excitations against a backdrop of topology.
By linking exciton condensation to the magnetic and topological degrees of freedom of MnBi$_2$Te$_4$, these results open avenues for magnetoelectric control of excitonic order and for leveraging axionic excitons in topological sensors and field-tunable quantum materials.
This framework provides a route toward electrically tunable exciton and polariton devices based on magnetic topological insulators, suggesting opportunities for dissipationless interlayer transport, exciton-based logic, and hybrid light--matter platforms for terahertz optoelectronics~\cite{Sun2019_ExcitonPolaritonTI, Ruta2023_VdWmagPolaritons, Basov2016_PolaritonsVDW, Mertz2024_ExcitonPolaritonsReview}.

\section{Low energy model for bilayer $\text{MnBi}_2 \text{Te}_4$}
We consider for simplicity a bilayer configuration of \MBT: an antiferromagnetic topological insulator whose manganese atoms host the spin moment, as well as being a vdW material, see Fig.~\ref{fig:MBT_crystal_structure_schematic}(a) for the layer-septuple crystal structure. Indeed, it has already been experimentally demonstrated to achieve atomically thin flakes of of \MBT ~\cite{Liu.2020, Deng.2020}. 
Each layer additionally contains spin $S=1/2$ electrons, with spin $\bm \sigma$ that move in the plane. The two layers are connected by an externally applied displacement field, denoted $D$, which allows for tunable charge separation between the upper and lower layers, see Fig.~\ref{fig:MBT_crystal_structure_schematic}(b) for a schematic of the proposed set-up. 

We use an effective $\mathbf{k}\cdot\mathbf{p}$ approach, where $\boldsymbol{k}$ is the crystal momentum and $\boldsymbol{p}$ is the quantum mechanical momentum operator, 
such that the Hamiltonian describing the single-particle physics is 
\begin{equation}\label{eq:single_ple_H}
    \hat{H}_0 = \sum_{\bf k} \Psi^{\dagger}_{\bf k} \left[ \mathcal{H}_0(\mathbf{k}) - \mu \right] \Psi_{\bf k},
\end{equation}
where $\Psi_{\bf k}$ is the second-quantized electronic spinor in spin/layer space, and $\mu$ is the chemical potential, which we set to zero without loss of generality.  We take a Bloch Hamiltonian of 
\begin{equation}\label{eq:BlochH}
\mathcal{H}_0(\mathbf{k}) = [ v_F (\boldsymbol{k} \times \boldsymbol{\sigma}) \cdot \hat{e}_z +D + \boldsymbol{N} \cdot \boldsymbol{\sigma} ] \tau_3 ,
\end{equation}
for the bilayer \MBT system, which captures the relevant physics~\cite{Sun.2023, Sun.2020}. Each term in this Hamiltonian represents a crucial physical aspect of the system. 
These terms are, respectively, a Rashba-type Dirac surface state on each layer with chirality fixed to the layer (implying the material has a broken mirror plane symmetry but preserves overall inversion);  $D$ represents an external bias or displacement field applied between the two layers, which serves as a critical tuning parameter for driving a chemical potential imbalance between the upper and lower layers; and $\boldsymbol{N} \cdot \boldsymbol{\sigma}$ incorporates the intrinsic Néel order of the antiferromagnet. 
This last term reflects the observed magnetic ground state where spins are ferromagnetically aligned within each layer and antiferromagnetically between layers.
The layers are indexed by the pseudospin operator $\tau_3 = \pm 1$ for the upper and lower layers, respectively.  

The displacement field $D$ is the control knob for charge imbalance: it drives a finite population of electrons on the upper layer, and a finite density of holes on the lower layer. 
There will be a resultant Coulomb attraction due to this charge imbalance.
In the case of the spatially separated bilayer, we distinguish between two different channels for the Coulomb interaction: interlayer and intralayer.
We introduce the layer-resolved charge densities, which in real-space read as
\begin{align}
    \delta \rho_U(\boldsymbol{r}) = \Psi^{\dagger}(\boldsymbol{r}) \hat{\Pi}_U  \Psi(\boldsymbol{r}) , \\
    \delta \rho_L(\boldsymbol{r}) =  \Psi^{\dagger}(\boldsymbol{r}) \hat{\Pi}_L \Psi(\boldsymbol{r}) .
\end{align}
Here the projection operators $\hat{\Pi}_{U/L}$ of the electronic spinor $\Psi(\boldsymbol{r})$  act to localize the charge onto a layer, so they count the number of electrons on each layer. 
To be explicit, this matrix is written as 
$ \Psi(\boldsymbol{r}) = \begin{pmatrix}
    \psi_U(\boldsymbol{r}) \\ \psi_L(\boldsymbol{r})
\end{pmatrix}$, 
where each element is a spinor. 
In this treatment we will consider only the interlayer Coulomb interactions, $V$, which can lead to very stable excitons due to the large degree of spatial separation between the electron and hole.

The interacting Hamiltonian due to Coulomb interaction is thus
\begin{equation}
    \hat{H}_{int} = V \int d^2r \text{   } \delta \rho_U(\boldsymbol{r}) \delta \rho_
    L(\boldsymbol{r}) ,
\end{equation}
where the Coulombic interaction $V$ is taken to be short-ranged and constant.
This interacting Hamiltonian is quartic in the fermionic creation and annihilation operators. The full Hamiltonian is the sum of the interacting and non-interacting terms, $H = H_0 + H_{int}$. While the single-particle Hamiltonian $H_0$ is linear and diagonalizable, the interacting term requires further attention, as we show in the next section.

\section{Exciton condensate within mean field theory}
We employ mean field theory to handle the many-body interaction term $H_{int}$ which couples the bilinear density operators from the upper and lower layers, producing a quartic fermionic term. Within mean field, this quartic interaction term 
gets replaced by an effective averaged field, which we denote $ \hat{\Delta}$, by employing the Hubbard–Stratonovich transformation in the particle–hole sector, which serves to decouple the interaction channel and reduce it to an effective single particle problem.  Detailed derivations are provided in Appendix \ref{sec:Appendix_ExOP_derivation}. 

We begin in the imaginary-time path integral formalism, which allows us to consider finite temperature effects, where the fermionic action reads
\begin{equation}
S[\bar{\psi},\psi] = \int_0^{\beta}  d^3x \left( \bar{\psi} \left[ \frac{\partial}{\partial \tau} + \mathcal{H}_0(\boldsymbol{k})  \right] \psi  \right) + S_{int} ,
\end{equation}
with the notation $\int d^3x = \int_0^{\beta} d\tau \int d^2r $. 
Here $\tau \in (0, \beta)$ is imaginary time, for $\beta = 1 / T$ inverse temperature. Note we use natural units in this work, where $c = \hbar = k_B = 1$. The interlayer density-density interaction is simply $S_{int} = \int d\tau  H_{int} $. 
One can show (see Appendix \ref{sec:Appendix_ExOP_derivation}) that the interacting term becomes decoupled using the Hubbard-Stratonovich transformation with the introduction of the new auxiliary complex matrix field, $\hat{\Delta}(x)$, as
\begin{equation}
     \hat{S}_{int}  = \int d^3x \text{Tr} \left[ \frac{\hat{\Delta}^{\dagger} \hat{\Delta}}{V}  \right]  + \bar{\psi}(x) \left[ \hat{\Delta} \tau^+ + \hat{\Delta} \tau^-  \right] \psi(x) .
\end{equation}
Note that $\tau^{\pm} = (\tau_1 \pm i \tau_2) / 2$ act as raising/lowering operators between layers.  
This new field can be physically interpreted as the bosonic exciton order parameter, given by the correlation function $\hat{\Delta} = V \langle \bar{\psi}^{U} \psi^L \rangle$; or in the form of a complex matrix field $\hat{\Delta} = \Delta_0 + \boldsymbol{\Delta}\cdot \boldsymbol{\sigma}$. Here $\Delta_0$ denotes the un-spin-polarized order parameter and $\boldsymbol{\Delta}\cdot\boldsymbol{\sigma}$ allows for spin-polarized excitons.

\begin{figure*}
\includegraphics[width=\textwidth]{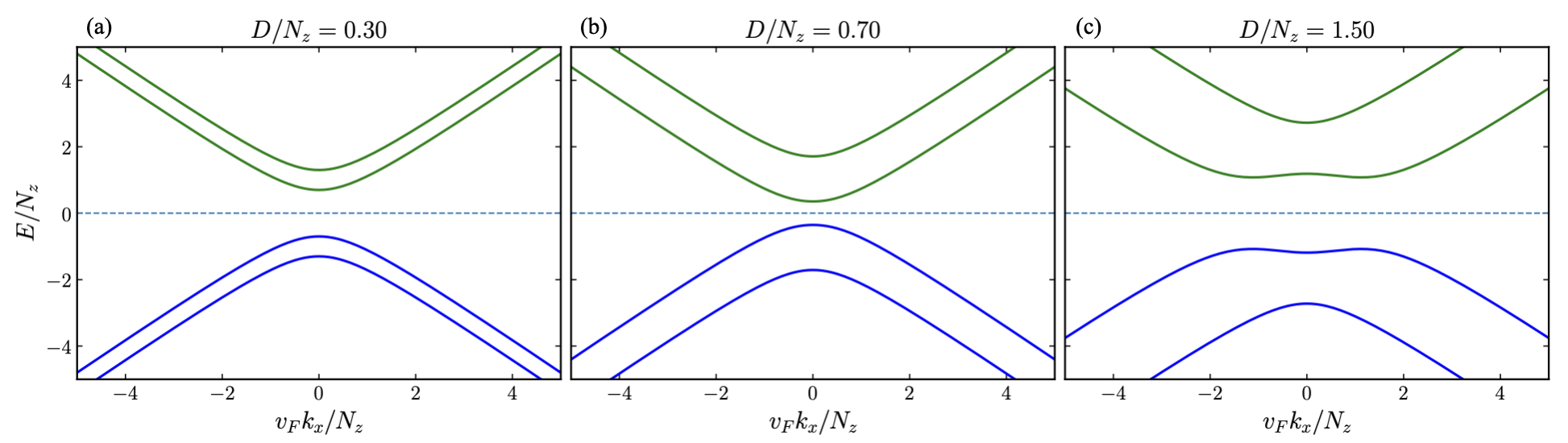}
\caption{\label{fig:BandStructures}Many-body band structure for constant $T/N_z = 0.15$ and $V N_z / v_F^2 = 0.285$. Band evolution as D is scanned across the mean field phase diagram from the uncondensed into the condensed phase. At small displacement field $D/N_z = 0.3$, the bilayer \MBT spectrum consists of two gapped Dirac cones separated by the Néel order $N_z$. As D increases through the condensate region of Fig.~\ref{fig:PhaseDiagram_SpectralFunction}(a)-(c), hybridization mediated by the exciton order parameter $\Delta_0$ drives a band inversion, signaling a possible topological phase transition. In this regime the condensate may be identified as an exciton insulator, where the excitonic gap supplants the single-particle Dirac mass. }
\end{figure*}

The exciton order parameter describes the coherent binding of an electron in the upper layer to a hole in the lower layer. To determine the conditions for the formation of a bilayer excitonic condensate, one may self-consistently solve for where $\hat{\Delta}$ is nonzero via its gap equation. This is obtained using the saddle point of the action with respect to $\hat{\Delta}$
\begin{equation}\label{eq:S_eff_minimize}
    \frac{\delta S_{\rm eff}}{\delta \hat{\Delta}} = 0,
\end{equation}
which yields the finite-temperature, self-consistent exciton gap equation as: 
\begin{equation}\label{eq:SCF_exciton}
    \frac{1}{V} \Delta^{\mu} = - \text{Tr} \frac{1}{\beta} \sum_{i \varepsilon_m} \int \frac{dp^2}{(2 \pi)^2}  \mathbb{G}[\hat{\Delta} ] \tau^- \sigma^{\mu},
\end{equation}
where $\mu=0,1,2,3$ labels the elements of the order parameter. The Matsubara Green's function is
\begin{align}\label{eq:G_action}
    \mathbb{G} =  \bigg[ i \varepsilon_m & - \left[  v_F (\boldsymbol{k} \times \boldsymbol{\sigma}) \cdot \hat{e}_z + D + N_z \sigma_z  \right] \tau^3 \cdots \nonumber \\
    & \cdots - \Delta^{\mu} \sigma^{\mu} \tau^+ - \bar{ \Delta}^{\mu} \sigma^{\mu} \tau^- \bigg]^{-1} ,
\end{align}
and $\varepsilon_m$ denotes the imaginary time, Matsubara frequencies. 


The mean-field phase diagram shown in Fig.~\ref{fig:PhaseDiagram_SpectralFunction}(a)-(c) illustrates the conditions for exciton condensation obtained by solving the self-consistent gap equations~\eqref{eq:S_eff_minimize} - \eqref{eq:G_action}. 
We assume the spin-unpolarized excitons will be the dominant species due to the simplicity of the model, and so we restrict our attention to this flavor going forwards. 
At low temperatures and moderate displacement field $D$, the exciton order parameter with no spin polarization, $\Delta_0$, acquires a finite value, marking the onset of a condensate. 
As temperature increases, thermal fluctuations suppress coherence, and the order parameter vanishes above a critical temperature threshold. Similarly, at small displacement fields the condensate is disfavored: as $D$ goes to zero, the charge imbalance between layers is insufficient to stabilize bound electron–hole pairs. This leads to a dome-like region of stability in the $(D,T)$ plane. Notably, the overall scale of $\Delta_0$ and the width of the condensate dome are set by the Néel order, which controls the underlying Dirac gap and thus the available phase space for exciton formation. These results highlight the strong interplay between displacement field, temperature, and magnetic order in tuning exciton condensation in bilayer \MBT.

\section{Obtaining the spectral function}
To compare theoretical modeling to experimental findings requires derivation of the exciton spectral function.
In order to characterize the exciton spectrum in the uncondensed phase ($T > T_c$ for $T_c$ temperature of exciton condensation), we expand the Matsubara action beyond mean field using the random-phase approximation (RPA). Integrating out the fermionic degrees of freedom yields the quadratic fluctuation effective action
\begin{equation}
    S = \sum_{q} \mathcal{L}_{\mu \nu}(i\Omega_m, q)  \Delta^{\mu}_q \bar{\Delta}^{\nu}_{q} ,
\end{equation}
where $\mathcal{L}_{\mu \nu}(i\Omega_m, q)$ is the inverse exciton propagator that encodes the collective-mode correlations of the auxiliary Hubbard-Stratonovich field $\hat{\Delta}$. This is given explicitly as 
\begin{align}
      \mathcal{L}_{\mu \nu}& (i \Omega_m, \boldsymbol{q})  = \frac{1}{V}\delta_{\mu \nu} \cdots \nonumber\\
      & \cdots + \frac{1}{2} \int_p {\rm Tr} \bigg( \hat{\mathbb{G}}_0 (p - q/2) \sigma^{\mu} \tau^- \hat{\mathbb{G}}_0 (p + q/2) \sigma^{\nu} \tau^+ \cdots \nonumber \\ 
     & \cdots + \hat{\mathbb{G}}_0 (p - q/2) \sigma^{\mu} \tau^+  \hat{\mathbb{G}}_0 (p + q/2)\sigma^{\nu} \tau^- \bigg) , \label{eq:Lmunu} 
\end{align}
with $\hat{\mathbb{G}}_0(p)$ the single-particle Green's function evaluated in the uncondensed state $\langle \hat{\Delta} \rangle = 0$ (see Appendix~\ref{Appendix:BMFT} for detailed derivation). Note $p = (\boldsymbol{p},i\varepsilon_m)$ label the internal fermionic degrees of freedom, where $\boldsymbol{p}$ are the fermion momentum and $\varepsilon_m = 2 \pi T (m + 1/2)$ is the fermionic Matsubara frequencies. Meanwhile $q = (\boldsymbol{q}, i \Omega_n)$ label the external bosonic collective modes, where $\boldsymbol{q}$ is the momentum carried by the exciton collective mode, and $\Omega_n$ are the bosonic Matsubara frequencies, which enter as arguments of the bosonic propagator. 

For the un-spin-polarized channel ($\mu = \nu = 0$) and $\boldsymbol{q} = 0$, equation~\eqref{eq:Lmunu} becomes 
\begin{align}
    & \mathcal{L}_{00}(i \Omega_n, 0) = \frac{1}{V} +  \int_{\boldsymbol{p}}   \frac{d^2p}{(2 \pi)^2}   \sum_{i\varepsilon_m} \times \cdots\label{eq:L00}  \\ 
    & \bigg(\frac{ (i \varepsilon_m -D) (i \varepsilon_m + D + i \Omega_n) - E_{\boldsymbol{p}}^2}{[(i \varepsilon_m - D )^2 - E_{\boldsymbol{p}}^2] [ (i \varepsilon_m + D + i \Omega_n)^2 - E_{\boldsymbol{p}}^2 ]}   + \{ D\rightarrow -D \}  \bigg) \nonumber , 
\end{align}
where $E_{\boldsymbol{p}} = \sqrt{N_z^2 + v_F^2 \boldsymbol{p}^2}$ is the gapped Dirac dispersion.

After performing the Matsubara-frequency summation using the residue theorem, analytic continuation $i \Omega_n \rightarrow \Omega + i 0^+$ gives the retarded Green's function response. The excitonic spectral function for $\boldsymbol{q}=0$ follows as 
\begin{equation}
   \mathcal{A}_{00}(\Omega,0) = - \frac{1}{\pi} \text{Im} \mathcal{L}_{00}^{-1}( \Omega, 0) . \label{eq:SpectralFcn}
\end{equation}
The peaks in the spectral function below the gap at $\Omega \simeq 2E_{\boldsymbol{p}}(k=0) = 2 N_z$ indicate an interlayer exciton collective mode: the position of the peak $\Omega(\boldsymbol{q})/N_z$ give the mode energy, while the width of the peak encodes the damping, or lifetime, of the mode.

As can be seen from the spectral function plots in Fig.~\ref{fig:PhaseDiagram_SpectralFunction}(d)-(f), the three variables under consideration are  V, D, and T. 
Lower values of V allow for collective mode formation and result in strong peaks alongside sufficiently low temperature and strong D, see Fig.~\ref{fig:PhaseDiagram_SpectralFunction}(d). Meanwhile higher V leads to collective mode softening and an instability towards condensation, Fig.~\ref{fig:PhaseDiagram_SpectralFunction}(f). 
The displacement field sets the interlayer detuning, i.e. the relative distance between conduction and valence bands between the two coupled layers, see Fig.~\ref{fig:BandStructures}. Before the condensate forms and D is increased, this brings the valence and conduction bands closer together, see Figs.~\ref{fig:BandStructures}(a) and (b), making it easier for excitons to form. Once a critical threshold bias field is reached, the excitons condense and the bands invert, see Fig.~\ref{fig:BandStructures}(c). 
A lower D means it takes more energy to couple electrons to holes, resulting in shorter lifetimes and exciton peaks nearest the gap. A larger D field softens the modes as it takes increasingly less energy to form bound pairs. For higher D, thermal fluctuations play less of a role as the strong chemical potential imbalance drives electrons and holes to become correlated. As the D field is lowered, thermal fluctuations play a more significant role to block exciton formation and drive the exciton peak towards the particle-hole continuum, see Fig.~\ref{fig:PhaseDiagram_SpectralFunction}(f). 

The exciton spectral function is the bridge between theoretical prediction and experimental observation. 
Experimentally, several probes are used to directly measure this quantity. Momentum-resolved spectroscopies such as resonant inelastic X-ray scattering (RIXS)~\cite{Ament.2011} or electron energy-loss spectroscopy (EELS)~\cite{Colliex.2022} measure intensities proportional to $\mathcal{A}_{\mu \nu}(i\Omega_m,q)$. 
Optical absorption or reflectivity experiments~\cite{Ridolfi.2018} probe the $q=0$ component of the same spectral weight, while photoluminescence measurements~\cite{Brem.2020} detect radiative recombination of excitons, again reflecting features in the spectral function.

Thus, the theoretical calculations of $\mathcal{A}_{\mu \nu}(i\Omega_m,\boldsymbol{q})$ provide a direct link to observables: sharp excitonic resonances predicted below the particle–hole continuum correspond to well-defined peaks in RIXS or optical spectra. By comparing the dispersion and linewidth of these peaks across theory and experiment, one can establish signatures of exciton formation and assess the stability of the condensate in bilayer \MBT. 


\section{Conclusion and Discussion}
Our results demonstrate that a bilayer configuration of the antiferromagnetic topological insulator \MBT can host a tunable exciton condensate whose properties are strongly governed by the interplay between displacement field, Néel order, and Coulomb interaction (related to critical temperature). Within mean-field theory, we identify a dome-shaped region in the (D,T) plane where the non-spin-polarized order parameter $\Delta_0$ becomes finite, indicating a stable condensate driven by interlayer Coulomb attraction. Scanning across this phase diagram modifies the underlying band structure: at small $D$, the system remains a trivial gapped Dirac insulator, while at intermediate fields exciton-mediated hybridization produces a band inversion as shown in Fig.~\ref{fig:BandStructures}(c). This signals a transition into an exciton-insulator phase, in which the excitonic gap replaces the single-particle Dirac mass.

Beyond mean-field, our RPA analysis of the spectral function $\mathcal{A}_{00}(\Omega_m, \boldsymbol{q}) $ reveals the emergence of sharp excitonic resonances in the uncondensed phase. As $D$ approaches the condensation threshold, these modes soften, indicating the instability toward spontaneous coherence, while at higher $D$ the exciton resonance hardens and broadens as the bound state dissolves into the particle–hole continuum. The spectral peaks predicted here correspond directly to experimental observables accessible through RIXS, EELS, and optical spectroscopy, providing measurable fingerprints of exciton formation and collective-mode dynamics in \MBT bilayers.

More broadly, these findings place the bilayer configuration of \MBT in a unique class of correlated magnetic topological materials, where collective exciton order coexists and interacts with the system’s intrinsic axion electrodynamics. Because the axion angle $\theta$ in \MBT can vary dynamically in response to $\boldsymbol{E}\cdot\boldsymbol{B}$ drive, one could investigate in future work the coupling of the exciton and axion degrees of freedom. 
This indicates a regime where exciton condensation and axion electrodynamics become intertwined, such that the excitonic gap both reflects and renormalizes the material’s topological magnetoelectric response.

Our theoretical framework establishes the microscopic foundations for these phenomena and delineates the conditions under which they may emerge. By demonstrating how excitonic order can be tuned through an external bias, this work deepens the understanding of correlated bosonic topological states in magnetically ordered topological insulators, offering new routes toward field-tunable topological functionality.
In the long term, these insights may inspire novel functionalities in gate-tunable optoelectronics~\cite{Mak.2016,Xiong.2023}, where exciton coherence and axion-mediated magnetoelectric effects could be harnessed to engineer non-reciprocal photonic responses~\cite{Sun2019_ExcitonPolaritonTI, Ruta2023_VdWmagPolaritons, Basov.2016} or coherent interlayer tunneling devices~\cite{Wang.2023, Eisenstein2004, Qi2025}.  Moreover, exploiting these low-energy collective modes for novel solid-state dark matter detectors which rely on the formation of axionic exciton-polaritons could offer a pathway to ultralight dark matter detection~\cite{Marsh.2019, Qiu.2025, PhysRevD.102.095005}

\begin{acknowledgments}
The authors would like to acknowledge fruitful discussions with Dr. Ioannis Petrides, Dr. Zexun Lin, and Jonathan Sanchez-Lopez.
This work was supported by the Quantum Science Center (QSC), a National Quantum Information Science Center of the U.S. Department of Energy (DOE). We also acknowledge Grant Numbers GBMF8048 and GBMF12976 from the Gordon and Betty Moore Foundation.
\end{acknowledgments}

\bibliography{references}
\bibliographystyle{unsrt}

\newpage

\onecolumngrid

\appendix
\section{Mean field and exciton order parameter}\label{sec:Appendix_ExOP_derivation}
Here we present a more detailed derivation of the exciton order parameter, starting from the interacting Hamiltonian:
\begin{equation}
        \hat{H}_{int} = V \int d^2r \text{   } \delta \rho_U(\boldsymbol{r}) \delta \rho_
    L(\boldsymbol{r}) .
\end{equation}
Which is recast in the path integral formulation of quantum mechanics as the following action 
\begin{equation}
    S_{int} = V  \int d^3x ~ \delta \rho_U(\boldsymbol{r}) \delta \rho_
    L(\boldsymbol{r}), 
\end{equation}
where again the notation used is $\int d^3x = \int d \tau \int d^2 r$. 
The matrix which comprises the spin and layer resolved charge density operators is $\Psi(\boldsymbol{r}) =  \begin{pmatrix}
        \psi_U(\boldsymbol{r}) \\ \psi_L(\boldsymbol{r})
    \end{pmatrix}$, 
and its conjugate $\bar{\Psi}(\boldsymbol{r}) =  \begin{pmatrix} \bar{\psi}_U(\boldsymbol{r}) & \bar{\psi}_L (\boldsymbol{r})
\end{pmatrix}$. 
Each element of the upper and lower operators denote spin.  Explicitly, this is written as the annihilation operator
$\psi_U(\boldsymbol{r}) =  \begin{pmatrix}
        \psi_{U, \uparrow}(\boldsymbol{r}) \\ \psi_{U,\downarrow} (\boldsymbol{r})
\end{pmatrix}$ 
for the upper layer; 
while the creation operator on the lower layer 
$ \bar{\psi}_L =  \begin{pmatrix}
        \bar{\psi}_{L, \uparrow}(\boldsymbol{r}) & \bar{\psi}_{L, \downarrow} (\boldsymbol{r})
\end{pmatrix}$.

\vspace{2.5mm}

One may rewrite the $S_{int}$ in terms of the upper and lower spinors as
\begin{align}
    S_{int} = V  \int d^3x ~ \bar{\psi}_U (x) \psi_U (x)  \bar{\psi}_L (x) \psi_L(x)  = 
    - V \int d^3x ~\text{Tr} \left[ \psi_U(x) \bar{\psi}_L(x) \psi_L(x) \bar{\psi}_U(x) \right] .
\end{align}
The minus sign indicates that this is an attractive force: energy decreases if you increase the Coulomb potential. 
Now, defining the following complex matrices
\begin{equation}
    \Phi(x) = \psi_U(x) \bar{\psi}_L(x), \quad  \Phi^{\dagger}(x) = \psi_L(x) \bar{\psi}_U(x),
\end{equation}
gives the action 
\begin{equation}
    S_{int} = - V \int d^3x ~\text{Tr} \left[ \Phi(x) \Phi^{\dagger}(x)  \right] .
\end{equation}
Note, the product $\Phi(x) \Phi^{\dagger}(x) $ is always positive semi-definite, which in turn ensures the Coulomb interaction will always be attractive. 

\vspace{2.5mm}

Next, we perform the Hubbard-Stratonovich transformation with the introduction of a new, complex field $\hat{\Delta}$, which allows us to transform the interaction action with quadratic dependence on $\Phi(x)$ to a simpler form with linear dependence on $\hat{\Delta}$. This gives the following expression for $S_{int}$, now in terms of this new field 
\begin{equation}
     S_{int}  = - \int d^3x ~ \text{Tr} \left[ \frac{\hat{\Delta}^{\dagger} \hat{\Delta}}{V}  \right]  + \bar{\psi}(x) \left( \hat{\Delta} \tau^+ + \hat{\Delta} \tau^-  \right) \psi(x) .
\end{equation}
which in the form of complex fields may be expressed as $\hat{\Delta} = \Delta_0 + \boldsymbol{\Delta} \cdot \boldsymbol{\sigma}$ where $\boldsymbol{\sigma}$ denotes the triplet of Pauli spin matrices, allowing for the possibility of spin-polarized excitons. 

\vspace{2.5mm}

The full action is now given as 
\begin{align}
    & S = \int d^3x ~ \bar{\psi}(x)   \left( \frac{\partial}{\partial \tau} + \mathcal{H}_0(\boldsymbol{k}) \right) \psi(x) +  
     \int d^3x ~ \bar{\psi}(x) \left[ \hat{\Delta} \tau^+ + \hat{\Delta}^{\dagger} \tau^-  \right] \psi(x) + \frac{1}{V} \int d^3x ~\text{Tr} \left[ \hat{\Delta}^{\dagger} \hat{\Delta}  \right] \nonumber   \\
    & = \frac{1}{V} \int d^3x ~\text{Tr} \left[ \hat{\Delta}^{\dagger} \hat{\Delta}  \right] - \text{Tr}~ \text{ln}~\mathbb{G}^{-1}[\hat{\Delta}] \label{eq:Appendix_action_greensfnc} ,
\end{align}
where we've introduced the Matsubara Green's function $\mathbb{G}^{-1} = [i \varepsilon_m - \mathcal{H}_0(\boldsymbol{p}) - \hat{\Delta}\tau^+ - \hat{\Delta}\tau^-]$. Note $\varepsilon_m$ denotes the imaginary time, fermionic Matsubara frequencies. 

\vspace{2.5mm}

To see where the exciton order parameter is nonzero and obtain the mean-field phase diagram, we use the saddle point of the action with respect to $\hat{\Delta}$,
\begin{equation}
    \frac{\delta S_{\rm eff}}{\delta \hat{\Delta}} = 0,
\end{equation}
which yields
\begin{equation}\label{eq:Appendix_SCF_exciton}
    \frac{1}{V} \Delta^{\mu} = - \text{Tr} \frac{1}{\beta} \sum_{i \varepsilon_m} \int \frac{dp^2}{(2 \pi)^2}  \mathbb{G}[\hat{\Delta} ] \tau^- \sigma^{\mu},
\end{equation}
where $\mu=0,1,2,3$ label the elements of the order parameter, and  the Matsubara Green's function is
\begin{equation}
    \mathbb{G} =  \left[ i \varepsilon_m - \left[  v_F (\boldsymbol{p} \times \boldsymbol{\sigma}) \cdot \hat{e}_z + D + N_z \sigma_z  \right] \tau^3  - \Delta^{\mu} \sigma^{\mu} \tau^+ - \bar{\Delta}^{\mu} \sigma^{\mu} \tau^- \right]^{-1} .
\end{equation}
A condensate corresponds to a nonzero order parameter. The phase diagrams shown in Fig.~\ref{fig:PhaseDiagram_SpectralFunction}(a)-(c) plot the non-spin-polarized $\Delta_0$ order parameter as a function of displacement field $D$ and temperature $T$, at various Coulomb potentials $V$, in units normalized by the Néel order $N_z$. 

\section{Beyond mean field}\label{Appendix:BMFT}

Since the phase diagram of the exciton order parameter has large regions of the uncondensed phase, we now turn to study this phase where the order parameter is zero. To start, we rewrite the excitonic bilayer model with mean-field Hamiltonian 
\begin{align}
    & \hat{H} = \sum_{\boldsymbol{p}} \Psi^{\dagger}_{\boldsymbol{p}} \left[ \mathcal{H}(\boldsymbol{p}) \right] \Psi_{\boldsymbol{p}}, \\
    & \mathcal{H}(\mathbf{p}) = [ v_F (\boldsymbol{p} \times \boldsymbol{\sigma}) \cdot \hat{e}_z +D + \boldsymbol{N} \cdot \boldsymbol{\sigma} ] \tau_3  + \Delta^{\mu} \sigma^{\mu} \tau^+ + \bar{\Delta}^{\mu} \sigma^{\mu} \tau^- , \nonumber
\end{align}
where $\Delta^{\mu}$ are the excitonic orders derived self-consistently.

\vspace{2.5mm}

We want to derive the exciton spectrum for $T > T_c$ where $T_c$ is the temperature of equilibrium excitonic condensation.
This is found within the random phase approximation (RPA) from the Matsubara action after integrating out the fermions: 
\begin{align}
    & S = \sum_{q} \frac{1}{V} |\Delta^{\mu}_q|^2 - \text{Tr} ~ \text{ln}~  \hat{\mathbb{G}}^{-1}(p), \quad \text{where} \quad \hat{\mathbb{G}}^{-1}(p) = i \varepsilon_m - \mathcal{H}(\boldsymbol{p}) .
\end{align}
Note, $p = (\boldsymbol{p},i\varepsilon_m)$ label the internal fermionic degrees of freedom, where $\boldsymbol{p}$ is the fermion momentum and $\varepsilon_m = 2 \pi T (m + 1/2), ~m\in\mathbb{Z}$ are the fermionic Matsubara frequencies. Meanwhile $q = (\boldsymbol{q}, i \Omega_m)$ label the external bosonic collective modes, where $\boldsymbol{q}$ is the momentum carried by the exciton collective mode, and $\Omega_m$ are the bosonic Matsubara frequencies, which enter as arguments of the bosonic propagator. 

\vspace{2.5mm}

We may study fluctuations of the order parameter above $T_c$ assuming  $ \langle \Delta \rangle = 0$. We perform a Taylor series expansion around this averaged, zero value and apply many-body perturbation theory in the small fluctuation. Moreover,  $[\mathcal{H}_0(\boldsymbol{k}), \tau_3] = 0$, which amounts to diagonalizing  each layer separately to determine the coupling of fluctuations between layers. 
Substituting in the Green's function gives 
\begin{align}
   \text{Tr} ~\text{ln} \hat{\mathbb{G}}^{-1} = \text{Tr} ~\text{ln} \left [ \delta_{q,0} \hat{\mathbb{G}}^{-1}_0 (p) - 
\hat{\Delta} \tau^+ -  \hat{\Delta}^{\dagger}  \tau^-  \right] \\
    = \text{Tr} ~\text{ln} \left[ \mathds{1} - \hat{\mathbb{G}}_0 (p) \hat{\Delta}  \tau^+  - \hat{\mathbb{G}}_0  (p)\hat{\Delta}^{\dagger} 
 \tau^-  \right] , \label{eq:TrLnToExpand}
\end{align}
where $\hat{\Delta} = \Delta^{\mu} \sigma^{\mu}$ and $\hat{\Delta}^{\dagger} =  \bar{\Delta}^{\nu} \sigma^{\nu}$. We perform a double Taylor series expansion of Eq.~\eqref{eq:TrLnToExpand} for $\hat{\Delta}$ and $\hat{\Delta}^{\dagger} $ to second order, using expansion of the logarithm ${\rm ln}(1-x) \approx -x - \frac{x^2}{2} - \mathcal{O}(x^3) $. The lowest surviving order is the cross term
\begin{equation}
    \text{Tr}  \left[ \frac{1}{2} (\hat{\mathbb{G}}_0 (p) 
 \hat{\Delta} \tau^+) (\hat{\mathbb{G}}_0 (p)  \hat{\Delta}^{\dagger} \tau^- )\right] .
\end{equation}
The effective action becomes 
\begin{equation}
    S_{\rm eff} = \sum_{\mu, \varepsilon} \frac{1}{V} |\Delta^{\mu}|^2 + {\rm Tr} \left[ \frac{1}{2} (\hat{\mathbb{G}}_0 (p) 
 \hat{\Delta} \tau^+) (\hat{\mathbb{G}}_0 (p)  \hat{\Delta}^{\dagger} \tau^- )\right]  ,\label{eq:Appendix_EffAction}
\end{equation}
and we rewrite the expression above as 
\begin{equation}
     S = \sum_{q} \mathcal{L}_{\mu \nu}(q)  \Delta^{\mu}_q \bar{\Delta}^{\nu}_{q},
\end{equation}
with 
\begin{align}
      \mathcal{L}_{\mu \nu}( q)  = \frac{1}{V}\delta_{\mu \nu} &+ \frac{1}{2} \int_p {\rm Tr} \bigg[ \hat{\mathbb{G}}_0 (p - q/2) \sigma^{\mu} \tau^- \hat{\mathbb{G}}_0 (p + q/2) \sigma^{\nu} \tau^+  
     + \hat{\mathbb{G}}_0 (p - q/2) \sigma^{\mu} \tau^+  \hat{\mathbb{G}}_0 (p + q/2)\sigma^{\nu} \tau^- \bigg] .
\end{align}
So that the $\mathcal{L}_{\mu \nu}(q) $ matrix contains all the coefficients for the terms in Eq.~\eqref{eq:Appendix_EffAction}. Note the $\hat{\Delta}$ correlation functions are found as 
\begin{align}
    - \langle \Delta^{\mu}_q \bar{\Delta}^{\nu}_{q}  \rangle \equiv D^{\mu \nu}(q) \\
    \rightarrow D^{\mu \nu} (q) = - [\hat{ \mathcal{L}}(q) ]^{-1}_{\mu \nu}
\end{align}
Therefore we can obtain the spectral function $\mathcal{A}_{\mu \nu}(i\Omega_n,q)$ of $\hat{\Delta}$ by analytically continuing $\mathcal{L}_{\mu \nu}^{-1}(q)$.

\vspace{2.5mm}

This can be evaluated by noting that in $\hat{\mathbb{G}}_0 $, the exciton order parameter $ \hat{\Delta} = 0$ so that $[\hat{\mathbb{G}}_0 , \tau_3] = 0$ such that we can diagonalize $\hat{\mathbb{G}}_0 $ in each layer separately (in uncondensed phase $\hat{\mathbb{G}}_0 $ is diagonal in layer).  We diagonlize in each subpsace as:
\begin{equation}
    \hat{\mathbb{G}}_0^{-1}(i \varepsilon_m,\bm{p}) = i \varepsilon_m - \tau_3 \hat{h}_{\boldsymbol{p}} \quad \text{with} \quad \hat{h}_{\boldsymbol{p}} = D + \boldsymbol{N}\cdot\boldsymbol{\sigma} + v_F (\boldsymbol{p}\times \boldsymbol{\sigma})_z
\end{equation}
This gives the following expression for the $\mathcal{L}_{\mu \nu}$ matrix elements
\begin{align}
      & \mathcal{L}_{\mu \nu}(i \Omega_n, \boldsymbol{q})  = \frac{1}{V}\delta_{\mu \nu} + \frac{1}{2}   \int_p \text{Tr}  \bigg[   \hat{\mathbb{G}}_0^+(i \varepsilon_m, \boldsymbol{p}) \sigma^{\mu}  \hat{\mathbb{G}}_0^-(i \varepsilon_m + i \Omega_n, \boldsymbol{p}) \sigma^{\nu}   + \hat{\mathbb{G}}_0^- (i \varepsilon_m, \boldsymbol{p}) \sigma^{\mu} \hat{\mathbb{G}}_0^+ (i \varepsilon_m + i \Omega_n, \boldsymbol{p}) \sigma^{\nu} \bigg]
      \\
      & =  \frac{1}{V}\delta_{\mu \nu} + \frac{1}{2} \int_p \text{Tr}  \bigg[ \bigg(  \frac{[ i \varepsilon_m - D - \boldsymbol{N}\cdot\boldsymbol{\sigma} - v_F (\boldsymbol{p}\times \boldsymbol{\sigma})_z]}{(i \varepsilon_m - D)^2 - E^2_{\boldsymbol{p}}} 
      \sigma^{\mu}  \frac{[ i \varepsilon_m + i\Omega_n +  D + \boldsymbol{N}\cdot\boldsymbol{\sigma} + v_F (\boldsymbol{p}\times \boldsymbol{\sigma})_z]}{(i \varepsilon_m + i\Omega_n + D)^2 - E^2_{\boldsymbol{p}}} \sigma^{\nu} \bigg) \cdots \nonumber \\
      & ~~~~~~~~~~~~~~~~~~~~~~~~~~\cdots + \bigg(  \frac{[ i \varepsilon_m + D - \boldsymbol{N}\cdot\boldsymbol{\sigma} - v_F (\boldsymbol{p}\times \boldsymbol{\sigma})_z]}{(i \varepsilon_m + D)^2 - E^2_{\boldsymbol{p}}} 
      \sigma^{\mu}  \frac{[ i \varepsilon_m + i\Omega_n -  D + \boldsymbol{N}\cdot\boldsymbol{\sigma} + v_F (\boldsymbol{p}\times \boldsymbol{\sigma})_z]}{(i \varepsilon_m + i\Omega_n - D)^2 - E^2_{\boldsymbol{p}}} \sigma^{\nu} \bigg) \bigg]
\end{align}
We will only focus on the $q=0$, $\mu= \nu = 0$ un-spin-polarized case, which is motivated by the fact that this was the only non-zero order parameter: 
\begin{align}
    & \mathcal{L}_{00}(i \Omega_n, 0) = \frac{1}{V} +  \int_{\boldsymbol{p}}   \frac{d^2p}{(2 \pi)^2}   \sum_{i\varepsilon_m} \bigg(\frac{ (i \varepsilon_m -D) (i \varepsilon_m + D + i \Omega_n) - E_{\boldsymbol{p}}^2}{[(i \varepsilon_m - D )^2 - E_{\boldsymbol{p}}^2] [ (i \varepsilon_m + D + i \Omega_n)^2 - E_{\boldsymbol{p}}^2 ]}  \cdots \nonumber \\
    & ~~~~~~~~~~~~~~~~~~~~~~~~~~~~~~~~~~~~~~\cdots +  \frac{ (i \varepsilon_m + D) (i \varepsilon_m - D + i \Omega_n) - E_{\boldsymbol{p}}^2}{[(i \varepsilon_m + D )^2 - E_{\boldsymbol{p}}^2] [ (i \varepsilon_m - D + i \Omega_n)^2 - E_{\boldsymbol{p}}^2 ]}  \bigg) .
\end{align}
We solve for the poles of this matrix element via the residue theorem, and obtain the following expression: 
\begin{align}
& \mathcal{L}_{00}(i\Omega_n,0)
= \frac{1}{V}
- \int \!\frac{d^2 p}{(2\pi)^2}\,  \bigg[
\frac{f\!\big(\beta(D-E_{\mathbf p})\big)}{\,2D+i\Omega_n-2E_{\mathbf p}\,}  +\frac{f\!\big(\beta(D+E_{\mathbf p})\big)}{\,2D+i\Omega_n+2E_{\mathbf p}\,}
\bigg]  + \bigg[
\frac{f\!\big(\beta(-D-E_{\mathbf p})\big)}{\,-2D+i\Omega_n-2E_{\mathbf p}\,}  +\frac{f\!\big(\beta(-D+E_{\mathbf p})\big)}{\,-2D+i\Omega_n+2E_{\mathbf p}\,}
\bigg] . \label{eq:lastone}
\end{align}
The notation used above defines the weighting function as $f(\beta z)=\tfrac12\tanh\!\Big(\frac{\beta z}{2}\Big)$. Eq.~\ref{eq:lastone} is then inverted to obtain the excitonic spectral function, whose poles correspond to the exciton frequencies. 

\vspace{2.5mm}

For the case for zero displacement field and in the uncondensed phase, the spectral function is a sharp quasiparticle peak such that it is a $\delta$-function in energy, with location in energy following the dispersion of the quasiparticle as  
$E_{\boldsymbol{p}} = \sqrt{N_z^2 + v_F^2 \boldsymbol{p}^2}$ for $v_F$ the Fermi velocity. Therefore, one only needs to integrate over momentum. 
The spectral function is solved by analytic continuation $i\Omega_n \rightarrow \Omega + i0^+$ 
\begin{equation}
    \mathcal{A}_{00}( \Omega, 0) = - \frac{1}{\pi} \text{Im} \left[ \mathcal{L}_{00}^{-1} ( \Omega, 0) \right] .
\end{equation}

\begin{figure}
\includegraphics[width=0.5\textwidth]{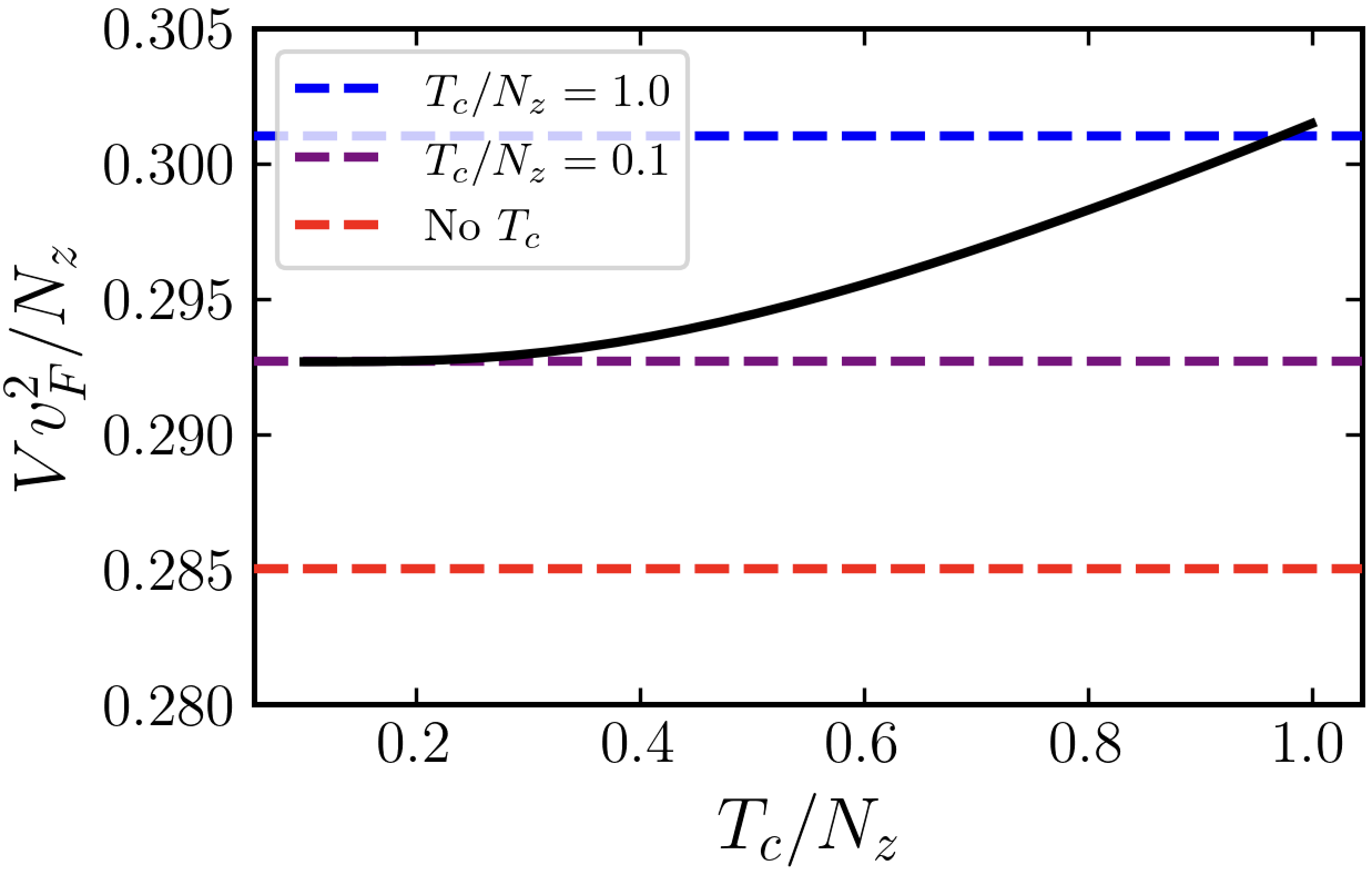}
\caption{\label{fig:Tc_vs_V} Dependence of the interlayer Coulomb interaction V on the critical exciton condensation temperature $T_c$. The solid black curve delineates where the self-consistent mean-field solution yields a finite exciton order parameter $\Delta_0$ at zero displacement field $D=0$. The red dashed line indicates an interaction strength V such that no exciton condensation can take place. The purple dashed line indicates a threshold voltage, defining the minimal interaction strength required to overcome thermal and single-particle energy scales for exciton binding. Physically, larger $V$ enhances interlayer electron–hole attraction, raising $T_c$ and stabilizing the condensate, whereas for $V < V_c$ only incoherent excitonic fluctuations persist. }
\end{figure}

\end{document}